\documentclass[letterpaper, 10 pt, conference]{ieeeconf}  % Comment this line out if you need a4paper

\IEEEoverridecommandlockouts                              % This command is only needed if 
\usepackage[utf8x]{inputenc}
\usepackage{hyperref}
\usepackage{graphicx}
\usepackage{wrapfig}
\usepackage{url}
\usepackage[export]{adjustbox}
\usepackage{physics} 
\usepackage{breqn}
\usepackage{hyperref}
\usepackage{amssymb}
\usepackage{amsmath}
\usepackage{siunitx}
\graphicspath{{images/}}

        \usepackage{float}
         \usepackage{epsfig} 
         \usepackage{graphicx}
         \usepackage{graphics}
         \usepackage{latexsym} 
         \usepackage{amsmath}
         \usepackage{psfrag}
         \usepackage{amsmath}
         \usepackage{amssymb}
         \usepackage{multicol}
         \usepackage{wrapfig}
         \usepackage{cite}
         \usepackage{url}
         \usepackage{bm} % bold math
        \usepackage{subcaption}
        
\usepackage{mathptmx}
\usepackage{anyfontsize}
\usepackage{t1enc}
        
\usepackage[usenames, dvipsnames]{color}        
\title{\LARGE \bf
A Lagrangian Model to Predict Microscallop Motion in non Newtonian Fluids
}

\author{Yashaswini Murthy$^{1}$ and Ravi Banavar$^{2}$% <-this % stops a space
\thanks{$^{1}$Department of Mechanical Engineering, Indian Institute of Technology Bombay, India.}
\thanks{$^{2}$Department of Systems and Control Engineering, Indian Institute of Technology Bombay, India.
}}

\begin{document}

\maketitle
\thispagestyle{empty}
\pagestyle{empty}

%%%%%%%%%%%%%%%%%%%%%%%%%%%%%%%%%%%%%%%%%%%%%%%%%%%%%%%%%%%%%%%%%%%%%%%%%%%%%%%%
\begin{abstract}

The need to develop models to predict the motion of microrobots, or robots of a much smaller scale,
moving in fluids in a low Reynolds number regime, and in particular, in non Newtonian fluids, cannot be understated. The article develops
a Lagrangian based model for one such mechanism - a two-link mechanism termed a microscallop,
 moving in a low Reynolds number
environment in a non Newtonian fluid. The modelling proceeds through the conventional Lagrangian
construction for a two-link mechanism and then goes on to model the external fluid forces using
empirically based models for viscosity to complete the dynamic model. The derived model is then 
simulated for different initial conditions and key parameters of the non Newtonian fluid, and the 
results are 
corroborated with a few existing experimental results on a similar mechanism under identical 
conditions. Lastly, with a view to implementing control algorithms we explore accessibility of the system at certain configurations.

%The objective of the work is to study model based predictions of a scallop moving in a non Newtonian fluid in the low Reynolds regime. The paper has five sections. The first section provides the reader with a general introduction to the key applications of micro robotics in the biomedical industry. This is followed by an introduction to a few preliminaries in mathematical modelling of locomoting systems in low Reynolds number. This is followed by a fairly detailed presentation on fluid dynamics,
%which plays a crucial role in comprehending locomotion in such environments. We then present an elementary Lagrangian modelling framework to arrive at the equations of motion of an isolated scallop in a non Newtonian fluid. We conclude the report with the results describing the behaviour of the scallop, obtained from simulating the obtained equations of motion.

\end{abstract}

%%%%%%%%%%%%%%%%%%%%%%%%%%%%%%%%%%%%%%%%%%%%%%%%%%%%%%%%%%%%%%%%%%%%%%%%%%%%%%%%
\section{INTRODUCTION}

The ability to access small spaces inside the human body at low Reynolds number (LRN) has facilitated research into controllable micro and nanorobotics. Alongside its myriad applications\cite{pap2},\cite{pap3},\cite{pap5}, of particular interest to this work is drug delivery, which in itself is a topic with vast room for technological improvements\cite{pap1}. To aid in drug delivery and many other minimally invasive applications of micro robots it is important to understand the environment in which they operate. Since most of the fluids in the human body are of the non Newtonian kind, it is necessary to study the properties of such fluids in the LRN regime in order to model microrobots for the same. To this end, the next section discusses certain existing mathematical models which describe the locomotive behaviour of organisms in the regime of interest. 

\section{MATHEMATICAL MODELLING}

The study of locomotion of micro organisms in LRN can be broadly classified into two categories: Newtonian and non Newtonian. 
In order to be able to design and control these microrobots, it is necessary to mathematically model them and understand their dynamical behaviour subject to manipulation of certain parameters. The simplest microrobot capable of locomotion has two rigid links and a single degree of freedom. \cite{purc} is an early paper which presents a fundamental theorem, concerning the motion characteristics of a scallop-like structure in the LRN regime. Henceforth, all our discussions are focused on locomotion in LRN regime.

\subsection{Scallop Theorem}
 A body in the LRN regime experiences a greater magnitude of  viscous  forces as when compared to the inertial forces. 
Consequentially, motion at that instant is entirely determined by the forces exerted at that moment and by nothing in the past\cite{purc}.
\\ The Navier Stokes equation, after neglecting the inertia terms, represents such fluid behaviour, and is given by: 
\begin{equation}\label{1}
    \eta \nabla^2 \Vec{v} - \nabla p = 0 \hspace{10mm} \nabla. \Vec{v} = 0
\end{equation}
The above equation is linear in space, and time independent\cite{leal},\cite{lauga1}.
When applied to LRN locomotion, the linearity and time-independence of the Stokes equation of motion lead to two important properties.
\begin{itemize}
    \item The first is rate independence: if a body undergoes surface deformation, the distance travelled by the swimmer between two different surface configurations does not depend on the rate at which the surface deformation occurs but only on its geometry (i.e. the sequence of shapes the swimmer passes through between these two configurations)\cite{lauga2}.
\end{itemize}
\begin{figure}[h]
\includegraphics[scale=.2]{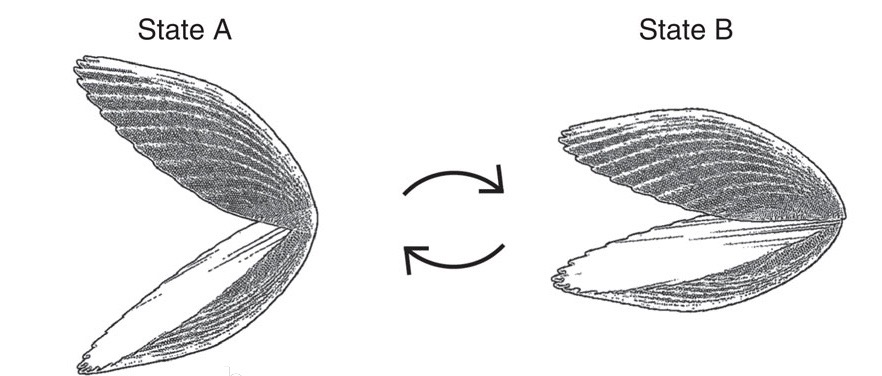}
\centering
\caption{\footnotesize{The open and close configurations of a scallop\cite{fish}}}  
\label{fig1}
\end{figure}
\begin{itemize}
    \item  The second important property is the so-called scallop theorem: if the sequence of shapes displayed by a swimmer deforming in a time periodic fashion is identical when viewed after a time-reversal transformation, then the swimmer experiences no motion on an average. 
    The only constraint is on the sequence of configurations of the swimmer and not the rate at which it executes its motion. 
    %Note that the condition is not that the motion be strictly time-reversal invariant, with the same forward and backwards rate, but only that the sequence of shapes is the same when viewed forward or backward in time. 
    This class of surface deformations is termed 'reciprocal deformation'. The scallop theorem introduces a strong geometrical constraint on the type of swimming motion which is effective at low Reynolds numbers\cite{lauga2}.
\end{itemize}
Various theories are used to model fluid bodies in the LRN regime. Two amongst these, that are the most commonly employed are the slender body theory\cite{slend} and the resistive force theory\cite{lauga3}. These will be discussed in the following section.

\section{FLUID DYNAMICS AND RESISTIVE FORCE THEORY}

Although non Newtonian fluids can be categorized into various categories based on their shear stress-strain rate behaviour, this paper primarily studies the time independent shear thinning non Newtonian fluids. 
\\
A shear thinning fluid is characterized by an apparent viscosity $\eta$ (defined as $\sigma_{xy}/\Dot{\gamma}_{xy}$, where $\sigma_{xy}$ is the shear stress and $\Dot{\gamma}_{xy}$ is the shear strain rate the fluid is subjected to) which gradually decreases with increasing shear rate. In polymeric systems (melts and solutions), at low shear rates, the apparent viscosity approaches a Newtonian plateau, where the viscosity is independent of shear rate (zero shear viscosity, $\eta_0$), given by: \\
\begin{equation}\label{2}
    \lim_{\Dot{\gamma}_{xy}\to 0}\frac{\sigma_{xy}}{\Dot{\gamma}_{xy}} = \eta_0 
\end{equation}
Furthermore, strictly polymer solutions also exhibit a similar plateau at very high shear rates (infinite shear viscosity, $\eta_\infty$), i.e.,
\begin{equation}\label{3}
     \lim_{\Dot{\gamma}_{xy}\to\infty}\frac{\sigma_{xy}}{\Dot{\gamma}_{xy}} = \eta_\infty
\end{equation}

\par There are various empirically developed models to capture the behaviour of shear thinning fluids,
% of which the Carreau viscosity model is of interest to us.\ 
a few of which are discussed below:\\

%\textit{Viscosity Models of Shear Thinning Fluids}
\begin{itemize}
    \item Power law model states that the relationship between shear stress($\sigma$) and shear rate($\Dot{\gamma}$) plotted on log-log co-ordinates for a shear-thinning fluid can be approximated by a straight line over an interval of shear rate as,
\begin{equation}\label{4}
    \sigma = m(\Dot{\gamma})^n
\end{equation}
where, $n$ is the power index and $m$ is a flow consistency index, both of which are characteristics of the fluid.
% or, in terms of the apparent viscosity,
% \begin{equation*}
%     \eta = m(\Dot{\gamma})^{n-1}
% \end{equation*}
% This law does not help us capture the aysmptotic values of viscosity at very low and very high shear rates.
\end{itemize}

% \begin{itemize}
%     \item Cross Viscosity Equation is obtained by correcting the power law to obtain the asymptotic limits of viscosity. 
% \begin{equation*}
%     \frac{\eta-\eta_\infty}{\eta_\infty-\eta_0} = \frac{1}{1 +  m(\Dot{\gamma})^n}
% \end{equation*}

% In the above two laws, obviously $0<n<1$, to have a shear thinning behaviour (viscosity decreasing with increasing shear rate). $n$ is an indication of the extent of shear thinning nature of the material while $m$ is an indication of the consistency of the material.
% \end{itemize}

\begin{itemize}
    \item Carreau viscosity model is another model which has been found to be more accurate with respect to the experimental results. The viscosity relationship according to such a model is presented below.
\begin{equation}\label{5}
    \mu_{eff}(\Dot{\gamma})=\mu_{inf}+(\mu_0-\mu_{inf})(1+(\lambda\Dot{\gamma})^2)^{\frac{n-1}{2}}
\end{equation}
where, \\
$\mu_0$ is viscosity at shear rate tending to zero\\
$\mu_{inf}$ is viscosity at shear rate tending to infinity\\
$\lambda$ is the relaxation time\\
%$n$ is power index\\ 
In the above two laws, it is necessary that $0<n<1$, to have a shear thinning behaviour.
At low shear rate, a Carreau fluid behaves as a Newtonian fluid, while at higher shear rate as a power law fluid\cite{carr}.
\end{itemize}

%These distinguishing properties of non Newtonian fluids are exploited by microswimmers to overcome the scallop theorem, enabling non zero displacement with reciprocal motion. A more detailed insight into the same is given below. 
Having understood the distinguishing properties of non Newtonian fluids, there are certain aspects of these fluids that can be exploited to overcome the scallop theorem and have non zero displacement with reciprocal motion.
Since coefficient of viscosity in a NN fluid is no longer a constant (it depends on the shear strain rate), the reciprocal theorem is no longer rate independent. The viscous resistive forces acting on the acting on the body will depend on its velocity, in both the backward and forward strokes. Different stroke velocities, lead to different forces in the two directions, yielding an overall nonzero displacement post every periodic set of motions. This rate dependency can be exploited to design a periodic swimmer in a NN fluid. \cite{lauga1}  
\\
However, the scallop theorem holds only for a single scallop in a uniform flow field. Two scallops engaging in reciprocal motion, with non trivial phase differences, lead to a non uniform flow field between the two, causing the collapse of the scallop theorem. Hence there is no many-scallop theorem \cite{lauga1}.
In order to determine the displacement that these bodies undergo, it is necessary to determine the forces acting on them due to the fluid. To model the forces acting on bodies with high aspect ratio (similar to the most commonly prevalent biological specimens) several theories have been proposed, of which resistive force theory is the most commonly employed. 

\subsection{Resistive Force Theory}  

Resistive force theory is incorporated to determine viscous resistive forces on bodies whose radius of curvature is of a lower order of magnitude than its length. It provides us with the linear force density as a linear function of the local surface velocity of the filament. This force density is determined in the transverse and longitudinal direction of the filaments. Slender body theory is included at times, to obtain more accurate determination of these resistive forces. To extend this theory to non Newtonian fluids, it is necessary to be able to estimate drag coefficients in such a fluid. 
\par
\begin{figure}[H]
\centering
\includegraphics[width=6.5cm, height=2.5cm]{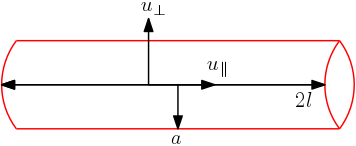}
\caption{\footnotesize{Straight filament of length $2l$ and cross-sectional radius $a$ may translate in a fluid along its length (velocity $u_\|$ ) or perpendicular to it ($u_{\bot}$).\cite{lauga3}}}    
\label{fig:pic2}
\end{figure}

 \cite{lauga3} proposes two approaches to calculate the non Newtonian drag coefficients. The first method involves deducing an empirical relation from experiments of sedimenting rods in shear-thinning fluids. The second method involves incorporating the Carreau visocity model to determine the drag coefficients. In order to determine these drag coefficients, it is necessary to determine the local shear strain rate at the surface of the filament, which is the same of the surface velocity (assuming no slip condition). We can then incorporate the shear-thinning nature of the fluid through a correction to the Newtonian drag coefficients and obtain a nonlinear velocity-force relationship. Much of our work on non-Newtonian fluid parameters and resistive force models is incorporated from the papers by Lauga \cite{lauga1}, \cite{lauga2} and \cite{lauga3}.\\
%In order to describe swimming induced by long, slender flagella, resistive force theory was proposed over 60 years ago and has subsequently been improved upon. For a radius of curvature of the flagellar waveform much larger than the diameter of the flagellum, then at leading order in the aspect ratio of the flagellum, the local velocity linearly determines the local force density on the flagellum. The drag can then be decomposed into the perpendicular and parallel components in this local region. Further refinements include slender-body theory, which provides greater accuracy leading to better qualitative and quantitative approximations. 
%\par  The first one is an empirical fit to experimental measurements of sedimenting rods in shear-thinning fluids and thus is directly built from experimental data. The second approach is an ad hoc model based on the Carreau viscosity-shear-rate relationship. Since in shear-thinning fluids the shear viscosity of the fluid is a function of the shear rate, we first need, in this case, a method to estimate accurately the local shear rate around the moving filament. %With this local, instantaneous value of the shear rate, 
Consider a filament of length $2l$ and radius $a$ as shown in Fig.\ref{fig:pic2}. 

The parallel and perpendicular drag coefficients in such a case assuming a locally Newtonian flow is given by\cite{lauga3}: 
\begin{equation}\label{6}
    \left.\frac{f_\|}{u_\|}\right\vert_{r=a} \equiv b_\| = \frac{4\pi\mu_0}{ln(\frac{4l^2}{a^2})-1}
\end{equation}
\begin{equation}\label{7}
    \left.\frac{f_\bot}{u_\bot}\right\vert_{r=a} \equiv b_\bot = \frac{8\pi\mu_0}{ln(\frac{4l^2}{a^2})+1}
\end{equation}
where, $f_\|$ and $f_\bot$ are the linear force densities in the parallel and perpendicular directions respectively. 
Upon further analysis it can be seen that the average shear rate due to both parallel and perpendicular motion is given by:
\begin{equation}\label{8}
    \Dot{\gamma}_{avg} = \frac{\sqrt{{f_\bot}^2 + 2{f_\|}^2}}{2\sqrt{2}a\pi\mu_0}
\end{equation}
where $\mu_0$ is the Newtonian viscosity of the medium\cite{lauga3}.

In case of non Newtonian fluids, $\mu_0$ is no longer a constant and is a function of strain rate. As a result, to propose drag coefficients to use with the Carreau model (or any emperical non Newtonian fluid model) we require the knowledge of the strain rates in the fluid near the filament. 
\\
Consider the Carreau fluid model, whose viscosity relationship is given by: 
\begin{equation*}
    \mu=\mu_{inf}+(\mu_0-\mu_{inf})(1+(\lambda\Dot{\gamma})^2)^{\frac{n-1}{2}}
\end{equation*}
%where, \\
% $\mu_0$ is viscosity at shear rate tending to zero\\
% $\mu_{inf}$ is viscosity at shear rate tending to infinity\\
% $\lambda$ is the relaxation time\\
% $n$ is power index\\ \\
Since high shear rates are unlikely to occur in our context, we set $\mu_{inf} = 0$, so the model simplifies to \begin{equation}\label{9}
    \frac{\mu}{\mu_0}=(1+(\lambda\Dot{\gamma})^2)^{\frac{n-1}{2}}
\end{equation} 
With the above simplification, we obtain the non Newtonian drag coefficients to be: 
\begin{equation} \label{10}
    b_{C_\bot} = R_C(u_\bot,u_\|)b_\bot, \hspace{5mm} b_{C_\|} = R_C(u_\bot,u_\|)b_\|
\end{equation}\\
where $b_\|$ and $b_\bot$ are given by \eqref{6} and \eqref{7} respectively and $R_C(u_\bot,u_\|)$ is the non Newtonian correction factor defined as below:
\begin{equation}\label{11}
    R_C = \bigg[1 + \bigg(\frac{\lambda b_\bot u_{\dot{\gamma}}}{2\sqrt{2}a\pi\mu_0}\bigg)^2\bigg]^{\frac{n-1}{2}},
\end{equation}\\
where $u_{\dot{\gamma}}$ is the shear rate velocity defined as:
\begin{equation}\label{12}
    u_{\dot{\gamma}} = \sqrt{{u_\bot}^2 + 2\frac{{b_\|}^2}{{b_\bot}^2}{u_\|}^2}
\end{equation}
where $u_\bot$ and $u_\|$ are the velocities of the fluid perpendicular and parallel to the slender filament\cite{lauga3}. 
With this preliminary introduction to fluid force modelling, we move onto the Lagrangian modelling of the scallop.

\section{Lagrangian Model of the Scallop}
\begin{center}
\begin{figure}[H]
   \centering
\includegraphics[width=7.5cm, height=5cm]{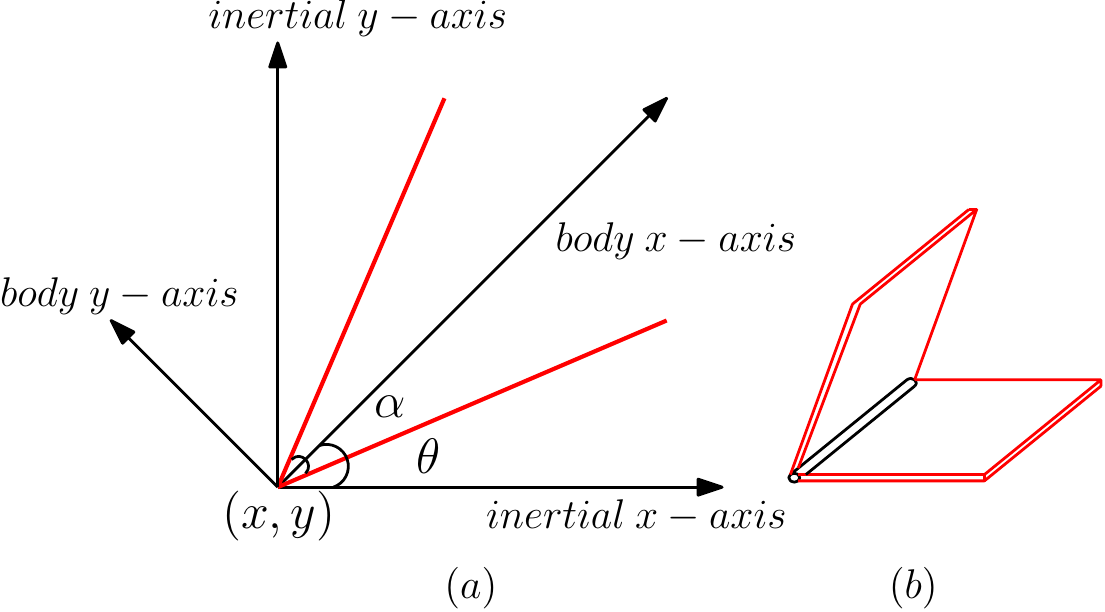}
\caption{\footnotesize{(a)The Schematic of a Scallop. (b)3 Dimensional model of a Scallop}}
\label{fig3}
\end{figure}
\end{center}

This section discusses the Lagrange model of the scallop in detail. Fig.\ref{fig3} presents the model of a scallop as two rigid links with a single hinge. 
\subsection{System Model} 
\vspace{-10mm}
The mass of the square links are $m_1$ and $m_2$, their lengths (as well as widths) are $L_1$ and $L_2$ and thicknesses are $a_1$ and $a_2$.
The model has four generalised coordinates, $(x,y,\theta,\alpha)$, $(x,y)$ representing the location of the hinge of the scallop with respect to a fixed coordinate axes, $\theta$, the angle between the scallop body axis and the inertial $x$ axis, and $\alpha$ being the angle between the scallop links. 
Since there are no conservative forces acting on the scallop, the Lagrangian will consist of only the kinetic energy due to the motion of the links. 
The force on the links of the scallop are calculated using resistive force theory as shown below.

\subsection{Forces on the links}

The force density, that is the force per unit area, acting on the links is taken as the drag coefficient times the strain rate (the velocity of the link in this case) and wave number\cite{lauga2},\cite{taylor}. The value of this coefficient varies according to the strain rate, since it is a non Newtonian fluid and is a medium property. The scallop can be assumed to be a rigid segment of a planar sheet whose wavelength is considerably larger than the scallop dimensions. The wave number accounts for the in-extensible planar sheet like structure of the scallop link, keeping the high aspect ratio of the link thickness and length intact. This serves as an extension to the resistive force theory employed in this context. %Since the opening and closing velocities of the scallop are different, we have different values for this coefficient, in the direction parallel to and perpendicular to the link. \\
The longitudinal drag coefficient is indicated by $b_{C_\|}$ and the lateral drag coefficient is by $b_{C_\bot}$ and are given by \eqref{10}.

The force density on link 1 is given by:  
\begin{equation}\label{13}
   F_{\parallel}(s)=b_{C_\|}(s)u_{\parallel}(s)k \hspace{5mm }F_{\bot}(s)=b_{C_\bot}(s)u_{\bot}(s)k,
\end{equation}
where $u_{\parallel}$ and $u_{\bot}$ represent the velocity parallel and perpendicular to the links.
\par Since $b_{C_\|}, b_{C_\bot}, v_{\parallel}, v_{\bot}$ all vary with distance along the length of the link, the forces also vary as a function of position along the length of the link.
 
The force densities acting on link 1 are found to be,
\begin{dmath}\label{14}
 F_{x1}(s)=F_{\parallel}(s)\cos(\theta + \frac{\alpha}{2})-F_{\bot}(s)\sin(\theta + \frac{\alpha}{2})
\end{dmath} 
\begin{dmath}\label{15}
 F_{y1}(s) = F_{\parallel}(s)\sin(\theta + \frac{\alpha}{2})+F_{\bot}(s)\cos(\theta + \frac{\alpha}{2})
\end{dmath}
Similarly the force density can be computed for the second link, with the only change being the angle made by the second link with the horizontal axis of the spatial frame.
The total force $f_{x1}$ in the $x$ direction and $f_{y1}$ in the $y$ direction on link 1 are given by integrating the force density along the length and width of the link, as shown:
\begin{equation}\label{16}
    f_{x1} = L_1\ast\int_{0}^{L_1}F_{x1}(s)ds, \hspace{5mm} f_{y1} = L_1\ast\int_{0}^{L_1}F_{y1}(s)ds
\end{equation}
Since we are assuming the force density to vary along the length and be independent of the width, the integration is with respect to the position along the length only. 
Similarly we can obtain the forces $f_{x2}$ and $f_{y2}$ acting on the second link.
\\ With the above computation, the Euler Lagrange Equations corresponding to the four generalised coordinates are derived.
% \vspace{-19mm}

\subsection{Euler Lagrange Equations}

%\vspace{-15mm}
% \setlength{\abovedisplayskip}{2pt}
\subsubsection{Equation of Motion corresponding to $x$}

\begin{equation} \label{17}
\vspace{-5mm}
\begin{split}
  f_{x1} + &f_{x2} = 
  \ddot{x}(m_1+m_2)-\frac{\dot{\theta}}{2}[m_1 L_1 \cos(\theta + \frac{\alpha}{2})(\dot{\theta} + \frac{\dot{\alpha}}{2})\\
    &+m_2 L_2 \cos(\theta- \frac{\alpha}{2})(\dot{\theta} - \frac{\dot{\alpha}}{2})]-
   \frac{\ddot{\theta}}{2}[m_1 L_1 \sin(\theta + \frac{\alpha}{2})\\  
   &+m_2 L_2 \sin(\theta- \frac{\alpha}{2})]-\frac{\dot{\alpha}}{4}[m_1 L_1 \cos(\theta + \frac{\alpha}{2})(\dot{\theta} + \frac{\dot{\alpha}}{2})\\
   &-m_2 L_2 \cos(\theta- \frac{\alpha}{2})(\dot{\theta}- \frac{\dot{\alpha}}{2})] -\frac{\ddot{\alpha}}{4}[m_1 L_1 \sin(\theta + \frac{\alpha}{2})\\
   &-m_2 L_2 \sin(\theta- \frac{\alpha}{2})]
\end{split}   
\end{equation}

where $f_{x1}$ and $f_{x2}$ are as described in the previous subsection.

\subsubsection{Equation of Motion corresponding to $y$}
\begin{equation}
\vspace{-2mm}
  \begin{split} \label{18}
  f_{y1} + &f_{y2} = \ddot{y}(m_1+m_2)-\frac{\dot{\theta}}{2}[m_1 L_1 \sin(\theta + \frac{\alpha}{2})(\dot{\theta} + \frac{\dot{\alpha}}{2})\\
  &+m_2 L_2 \sin(\theta- \frac{\alpha}{2})(\dot{\theta}- \frac{\dot{\alpha}}{2})]
  +\frac{\ddot{\theta}}{2}[m_1 L_1 \cos(\theta + \frac{\alpha}{2})\\
  &+m_2 L_2 \cos(\theta- \frac{\alpha}{2})]
  -\frac{\dot{\alpha}}{4}[m_1 L_1 \sin(\theta + \frac{\alpha}{2})(\dot{\theta} + \frac{\dot{\alpha}}{2})\\
  &-m_2 L_2 \sin(\theta- \frac{\alpha}{2})(\dot{\theta} - \frac{\dot{\alpha}}{2})]
   +\frac{\ddot{\alpha}}{4}[m_1 L_1 \cos(\theta + \frac{\alpha}{2})\\ &-m_2 L_2 \cos(\theta- \frac{\alpha}{2})]
\end{split}  
\end{equation}

where $f_{y1}$ and $f_{y2}$ are as described in the previous subsection.

\subsubsection{Equation of Motion corresponding to $\theta$}
The total torque acting on link 1 is given by:
\begin{equation} \label{19}
\tau_1=\int_{0}^{L_1} s_1*F_{\bot}(s)*L_1*ds_1 
\end{equation}
A similar expression could be obtained for $\tau_2$ acting on the second link. 
\begin{equation}\label{20}
 \begin{split} 
   \tau_1 + &\tau_2 = 
   +\frac{\dot{\theta}\dot{y}}{2}[m_1L_1\sin(\theta + \frac{\alpha}{2})+m_2L_2\sin(\theta - \frac{\alpha}{2})]\\
   &+\frac{\dot{\theta}\dot{x}}{2}[m_1 L_1 \cos(\theta + \frac{\alpha}{2})+m_2 L_2 \cos(\theta - \frac{\alpha}{2})]\\
   &+\frac{\dot{\alpha}\dot{y}}{4}[m_1L_1\sin(\theta + \frac{\alpha}{2})-m_2L_2\sin(\theta - \frac{\alpha}{2})]\\
   &+\frac{\dot{\alpha}\dot{x}}{4}[m_1 L_1 \cos(\theta + \frac{\alpha}{2})-m_2 L_2 \cos(\theta - \frac{\alpha}{2})]\\
   &+\frac{\ddot{\theta}}{3}[m_1L_1^2+m_2L_2^2]+\frac{\ddot{\alpha}}{6}[m_1L_1^2-m_2 L_2^2]\\&+ 
   \frac{\ddot{y}}{2}[m_1 L_1 \cos(\theta + \frac{\alpha}{2})+m_2 L_2 \cos(\theta- \frac{\alpha}{2})]\\&-\frac{\ddot{x}}{2}[m_1 L_1 \sin(\theta + \frac{\alpha}{2})+m_2 L_2 \sin(\theta - \frac{\alpha}{2})]\\ 
   &-\frac{\dot{y}}{2}[m_1 L_1 \sin(\theta + \frac{\alpha}{2})(\dot{\theta} + \dot{\frac{\alpha}{2}})\\&+m_2 L_2 \sin(\theta- \frac{\alpha}{2})(\dot{\theta} - \dot{\frac{\alpha}{2}})]\\&
  -\frac{\dot{x}}{2}[m_1 L_1 \cos(\theta + \frac{\alpha}{2})(\dot{\theta} + \dot{\frac{\alpha}{2}})\\&+m_2 L_2 \cos(\theta- \frac{\alpha}{2})(\dot{\theta} - \dot{\frac{\alpha}{2}})]
 \end{split}
\end{equation}

\subsubsection{Equation of Motion corresponding to $\alpha$}
\begin{equation} \label{21}
    \begin{split} 
   \frac{\tau_1 - \tau_2}{2} + &\tau= 
   +\frac{\dot{\theta}\dot{y}}{4}[m_1L_1\sin(\theta + \frac{\alpha}{2})-m_2L_2\sin(\theta - \frac{\alpha}{2})]\\
   &+\frac{\dot{\theta}\dot{x}}{4}[m_1 L_1 \cos(\theta + \frac{\alpha}{2})-m_2 L_2 \cos(\theta - \frac{\alpha}{2})]\\
   &+\frac{\dot{\alpha}\dot{y}}{8}[m_1L_1\sin(\theta + \frac{\alpha}{2})+m_2L_2\sin(\theta - \frac{\alpha}{2})]\\
   &+\frac{\dot{\alpha}\dot{x}}{8}[m_1 L_1 \cos(\theta + \frac{\alpha}{2})+m_2 L_2 \cos(\theta - \frac{\alpha}{2})]\\
   &+\frac{\ddot{\alpha}}{12}[m_1L_1^2+m_2 L_2^2]+ \frac{\ddot{\theta}}{6}[m_1L_1^2-m_2L_2^2]\\
   &+\frac{\ddot{y}}{4}[m_1 L_1 \cos(\theta + \frac{\alpha}{2})-m_2 L_2 \cos(\theta- \frac{\alpha}{2})]\\&-\frac{\ddot{x}}{4}[m_1 L_1 \sin(\theta + \frac{\alpha}{2})-m_2 L_2 \sin(\theta - \frac{\alpha}{2})]\\ 
   &-\frac{\dot{y}}{4}[m_1 L_1 \sin(\theta + \frac{\alpha}{2})(\dot{\theta} + \dot{\frac{\alpha}{2}})\\&-m_2 L_2 \sin(\theta- \frac{\alpha}{2})(\dot{\theta} - \dot{\frac{\alpha}{2}})]\\
  &-\frac{\dot{x}}{4}[m_1 L_1 \cos(\theta + \frac{\alpha}{2})(\dot{\theta} + \dot{\frac{\alpha}{2}})\\&-m_2 L_2 \cos(\theta- \frac{\alpha}{2})(\dot{\theta} - \dot{\frac{\alpha}{2}})]
\end{split}
\end{equation}

where $\tau$ is the externally applied torque for actuating the scallop links.

\section{Results and Discussion}

\subsection{Simulations}
For the purpose of conducting numerical simulations and corroborating with the existing experimental results, the physical characteristics of the scallop were taken as denoted in the following table. 
%\vspace{-3mm}
\begin{table}[H]
\caption{Microscallop Properties}
\centering
\begin{tabular}{|c || c |} 
 \hline
 Scallop Property & Value  \\  
 \hline\hline
 link length & $0.8mm$  \\ 
 \hline
 link breadth & $0.8mm$  \\
 \hline
 link thickness & $0.1mm$ \\
 \hline
 mass of single link & $61.76\ast10^{-9} kg$ \\ 
 \hline
 Wave number &  $1 m^{-1}$ \\ 
 \hline
\end{tabular}
\label{table:1}
\end{table}
\vspace{-3mm}
And the shear thinning fluid was taken to possess the following Carreau coefficients:
\vspace{-3mm}
\begin{table}[h!] 
\caption{Non Newtonian Fluid Properties}
\centering
\begin{tabular}{|c || c |} 
 \hline
 Fluid Property & Value  \\ 
 \hline\hline
Zero strain rate viscosity & $0.377 Pa-s$  \\ 
 \hline
Shear thinning coefficient ($n$) & $0.3$  \\
 \hline
 Relaxation time & $0.512s$ \\ 
 \hline
% Wave number &  $1 m^{-1}$ \\ 
% \hline
\end{tabular}
\label{table:2}
\end{table}
\vspace{-3mm}
\\The initial angle between the links is $\alpha(0) = 0.5618$ radians and the initial angle between the scallop body axis and the fixed horizontal axis is $\theta(0) = 0.523$ radians. 

Upon simulating the four equations, the following displacement trends were observed. In all but the last case, the torque applied has a time period of 4 seconds. 

\begin{enumerate}
    \item \textbf{Symmetric Actuation}: With identical opening and closing speeds there is negligible net displacement as expected. Since the resistive force depends on the strain rate, the force acting during the opening and closing of the scallop remains the same if the opening and closing strain rates are similar, thereby leading to no net displacement as depicted in Fig.\ref{fig5}-\ref{fig7}.
\\    
\begin{center}
\begin{figure}[H]
   \centering
\includegraphics[scale=.4]{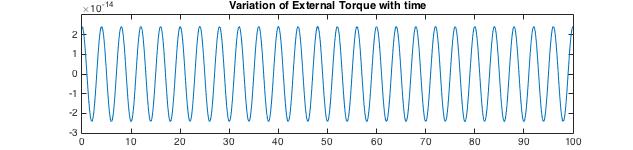}
\caption{\footnotesize{Variation of the External Torque acting on the scallop}}
\label{fig4}
\end{figure}
\vspace{-0.5cm}

\begin{figure}[H]
    \centering
\includegraphics[scale=.42]{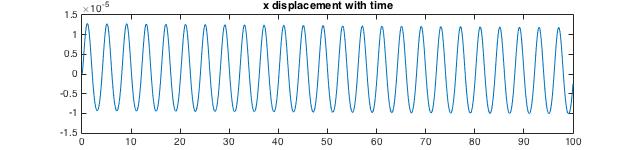} 
\caption{\footnotesize{Displacement in $x$ direction in $m/s$}}
\label{fig5}
\end{figure}
\vspace{-0.5cm}
\begin{figure}[H]
\includegraphics[scale=.42]{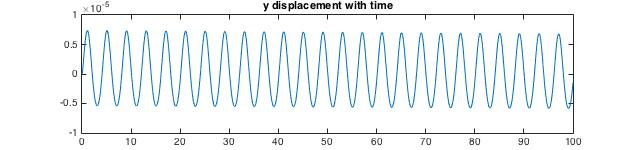}
\caption{\footnotesize{Displacement in $y$ direction in $m/s$}}
\label{fig6}
\end{figure}
 \vspace{-0.5cm}
%\caption{Displacement in metres against time in seconds}
\begin{figure}[H]
\includegraphics[scale=.42]{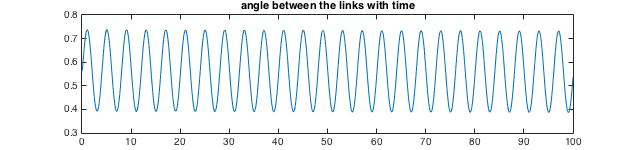}
\caption{\footnotesize{Variation of $\alpha$ in $rad/sec$}}
\label{fig7}
\end{figure}
% \vspace{-0.5cm}
% \begin{figure}[H]
% \includegraphics[scale=.42]{sym_th.jpg}
% \caption{Variation of $\theta$ in $rad/sec$}

% %\caption{Displacement in radians against time in seconds}
% \end{figure}
\end{center}
 
%\caption{Displacement in radians against time in seconds}
%\end{figure}
\vspace{-0.5cm}

\item \textbf{Asymmetric Actuation}: Upon asymmetric actuation of the scallop, with slow opening and fast closing of the links and the fluid properties as defined in Table \ref{table:2}, the displacement trends are as shown in Fig.\ref{fig9}-\ref{fig11}.  
\begin{center}
\begin{figure}[H]
   \centering
\includegraphics[scale=.4]{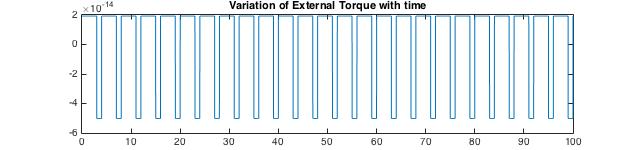}
\caption{\footnotesize{Variation of the External Torque acting on the scallop}}
\label{fig8}
\end{figure}
\vspace{-0.5cm}

\begin{figure}[H]
    \centering
\includegraphics[scale=.42]{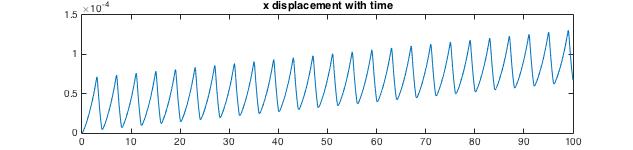} 
\caption{\footnotesize{Displacement in $x$ direction in $m/s$}}
\label{fig9}
\end{figure}
\vspace{-0.5cm}
\begin{figure}[H]
\includegraphics[scale=.42]{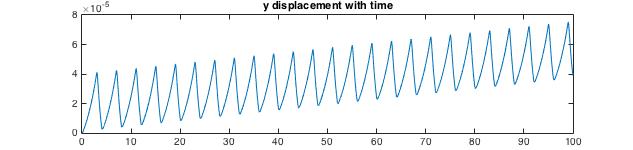}
\caption{\footnotesize{Displacement in $y$ direction in $m/s$}}
\label{fig10}
\end{figure}
 \vspace{-0.5cm}
%\caption{Displacement in metres against time in seconds}
\begin{figure}[H]
\includegraphics[scale=.42]{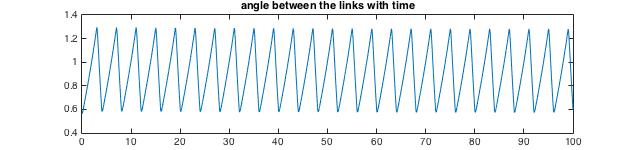}
\caption{\footnotesize{Variation of $\alpha$ in $rad/sec$}}
\label{fig11}
\end{figure}
%\vspace{-0.5cm}
% \begin{figure}[H]
% \includegraphics[scale=.42]{asym_th.jpg}
% \caption{Variation of $\theta$ in $rad/sec$}

% %\caption{Displacement in radians against time in seconds}
% \end{figure}
\end{center}
\vspace{-0.5cm}

\item \textbf{Higher zero shear rate viscosity}: Upon increasing the zero shear rate viscosity from $\mu_0 = 0.377$ to $\mu_0 = 0.6$, under asymmetric actuation as before, the resistive force increases accordingly leading to a decrease in the overall displacement to around $60 \mu m$ in $100s$ as can be observed in Fig.\ref{fig12}-\ref{fig14}.  
\begin{center}
\begin{figure}[H]
    \centering
\includegraphics[scale=.42]{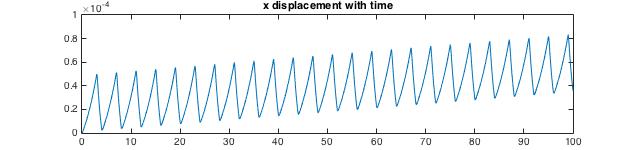} 
\caption{\footnotesize{Displacement in $x$ direction in $m/s$}}
\label{fig12}
\end{figure}
\vspace{-0.5cm}
\begin{figure}[H]
\includegraphics[scale=.42]{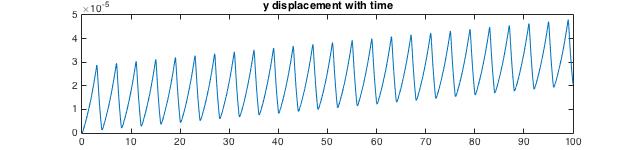}
\caption{\footnotesize{Displacement in $y$ direction in $m/s$}}
\label{fig13}
\end{figure}
 \vspace{-0.5cm}
%\caption{Displacement in metres against time in seconds}
\begin{figure}[H]
\includegraphics[scale=.42]{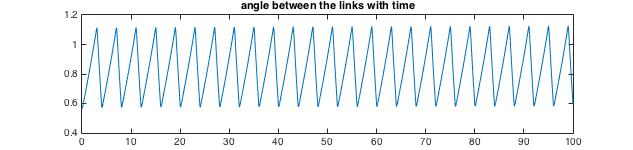}
\caption{\footnotesize{Variation of $\alpha$ in $rad/sec$}}
\label{fig14}
\end{figure}
% \vspace{-0.5cm}
% \begin{figure}[H]
% \includegraphics[scale=.42]{higheta_th.jpg}
% \caption{Variation of $\theta$ in $rad/sec$}

% %\caption{Displacement in radians against time in seconds}
% \end{figure}
\end{center}
\vspace{-0.5cm}

\item \textbf{Reduce shear thinning coefficient}: Upon decreasing the shear thinning coefficient from $n = 0.3$ to $n = 0.2$, under similar asymmetric actuation, the resistive force decreases (as the shear rate has a greater effect on the viscosity with decreasing shear thinning coefficient) accordingly leading to an increase in the overall displacement to around $200 \mu m$ in $100s$ as can be observed in Fig.\ref{fig15}-\ref{fig17}. 

\begin{center}
\begin{figure}[H]
    \centering
\includegraphics[scale=.42]{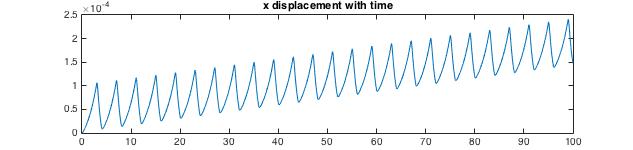} 
\caption{\footnotesize{Displacement in $x$ direction in $m/s$}}
\label{fig15}
\end{figure}
\vspace{-0.5cm}
\begin{figure}[H]
\includegraphics[scale=.42]{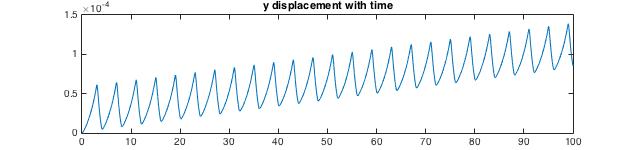}
\caption{\footnotesize{Displacement in $y$ direction in $m/s$}}
\label{fig16}
\end{figure}
 \vspace{-0.5cm}
%\caption{Displacement in metres against time in seconds}
\begin{figure}[H]
\includegraphics[scale=.42]{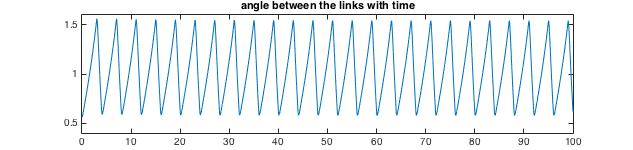}
\caption{\footnotesize{Variation of $\alpha$ in $rad/sec$}}
\label{fig17}
\end{figure}
% \vspace{-0.5cm}
% \begin{figure}[H]
% \includegraphics[scale=.42]{highn_th.jpg}
% \caption{Variation of $\theta$ in $rad/sec$}

% %\caption{Displacement in radians against time in seconds}
% \end{figure}
\end{center}
\vspace{-0.5cm}

\item \textbf{Increased time period}: When the time period of the actuation torque is increased from $4s$ to $6s$ as shown in Fig.\ref{fig18}, the displacement obtained is greater, as the opening and closing strokes have greater velocities and thereby a lesser resistive force. This behaviour is depicted in Fig.\ref{fig19}-\ref{fig21}, with the fluid properties as described in Table \ref{table:2}. 

\begin{center}
\begin{figure}[H]
   \centering
\includegraphics[scale=.4]{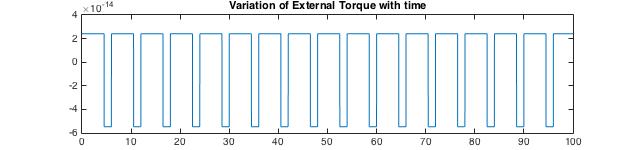}
\caption{\footnotesize{Variation of the External Torque acting on the scallop}}
\label{fig18}
\end{figure}
\vspace{-0.5cm}

\begin{figure}[H]
    \centering
\includegraphics[scale=.42]{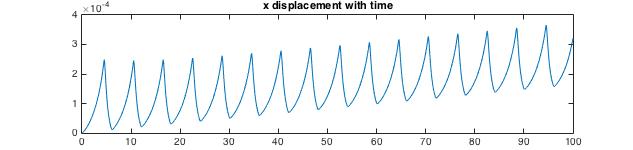} 
\caption{\footnotesize{Displacement in $x$ direction in $m/s$}}
\label{fig19}
\end{figure}
\vspace{-0.5cm}
\begin{figure}[H]
\includegraphics[scale=.42]{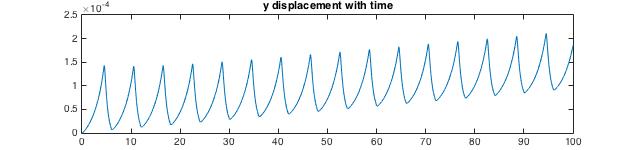}
\caption{\footnotesize{Displacement in $y$ direction in $m/s$}}
\label{fig20}
\end{figure}
 \vspace{-0.5cm}
%\caption{Displacement in metres against time in seconds}
\begin{figure}[H]
\includegraphics[scale=.42]{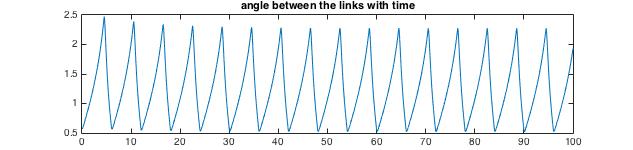}
\caption{\footnotesize{Variation of $\alpha$ in $rad/sec$}}
\label{fig21}
\end{figure}
% \vspace{-0.5cm}
% \begin{figure}[H]
% \includegraphics[scale=.42]{highper_th.jpg}
% \caption{Variation of $\theta$ in $rad/sec$}

% %\caption{Displacement in radians against time in seconds}
% \end{figure}
\end{center}
\end{enumerate}
\vspace{-0.5cm}

Since the external torque actuation is of the same magnitude (the only difference being the direction each link rotates in) for both the links, the scallop traverses in a straight line with the scallop axis at a constant angle with the inertial $x$ axis as shown in the figure below. 
%\vspace{-0.5cm}

\begin{figure}[H]
\includegraphics[scale=.42]{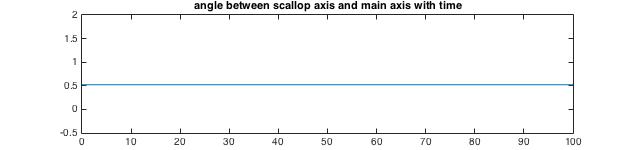}
\caption{\footnotesize{Variation of $\theta$ in $rad/sec$}}
%\caption{Displacement in radians against time in seconds}
\label{fig22}
\end{figure}

\par From the above simulation results, it is evident that the model adheres to the previously obtained trends in \cite{fish}. With this dynamic model at hand, we proceed to present the accessibility results of the microscallop. 

\subsection{Microscallop Accessibility}
An essential objective of the modelling is to explore control algorithms for motion planning
of the microscallop. In this subsection we explore a notion called accessibility at
certain configurations of the microscallop.  For the purpose of our simulations we have assumed the torques acting on the two links to be of equal magnitude and opposite in sign. But the two links can be subject to two different torques, giving rise to two control vector fields. Hence, the scallop system can be written as: 
\begin{equation}
    \dot{x}(t) = f_0(x(t)) + u^1(t)f_1(x(t)) + u^2(t)f_2(x(t))
\end{equation}
where $x(t)$ is a curve on an eight dimensional state manifold $M$ (corresponding to the four generalised coordinates and their velocities), $u^1$ and $u^2$ are the scalar torques acting on the two links and $\big\{u^1,u^2\} \in U$, where $U$ is a subset in $\mathbb{R}^2$. The vector field $f_0$ is the drift vector field, describing the dynamics of the system in the absence of controls, and the vector fields $f_1, f_2$ are the input vector fields or control vector fields, indicating how we are able to actuate the system.
For the above analytic control affine system, $\Sigma = (M,\mathcal{F}=\big\{ f_0,f_1,f_2 \} ,U)$, with $U$ proper, the accessibility theorem states that $\Sigma$ is accessible from $x$ if and only if $L(\mathcal{F})_{x} = T_{x}M$ \cite{lewis}.

\begin{figure}[H]
  \centering
  \begin{minipage}[b]{0.22\textwidth}
    \includegraphics[width=\textwidth]{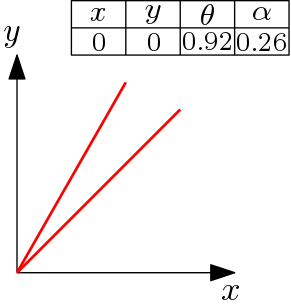}
  \end{minipage}
  \hfill
  \begin{minipage}[b]{0.22\textwidth}
    \includegraphics[width=\textwidth]{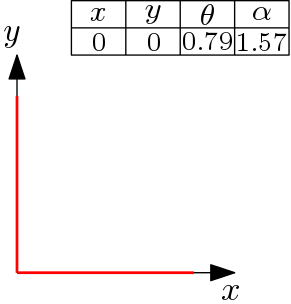}
  \end{minipage}
    \hfill
    \vspace{3mm}
  \begin{minipage}[b]{0.22\textwidth}
    \includegraphics[width=\textwidth]{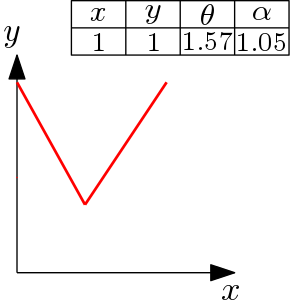}
  \end{minipage}
  \caption{Different scallop configurations at which accessibility was checked.}
  \label{fig:23}
\end{figure}

Accessibility was checked for the above positions of the scallop when it is in rest. The Lie brackets computed for this computation were $f_1, f_2, [f_0,f_1], [f_0,f_2], [f_0,[f_0,f_1]], [f_0,[f_0,f_2]], [f_1,[f_0,f_1]]$ and $[f_1,[f_0,[f_0,f_1]]]$, which spanned $T_xM$. Hence $\Sigma$ is accessible from these locations and other such configurations.

\section{CONCLUSION}
From the simulation figures we can infer that the velocity of the scallop along $\theta$ is nearly equal to $1-2 \mu m/s$. For a microscallop of similar dimensions and fluid properties, our technique, which is model based, predicts almost similar displacements as observed through experiments in \cite{fish}. Although we did conduct a non-dimensional analysis, it was necessary to resort to the current method in order to better understand the effect of fluid parameters on the scallop locomotion. The proposed control affine system is also accessible from various configurations. Hence, this model provides us with a means to generate other scallop motions through appropriately designed control laws.

%\bibliographystyle{unsrt}%Used BibTeX style is unsrt
%\bibliography{root}

\end{document}